\documentclass[oribibl]{llncs}
\usepackage{multicol}
\usepackage{amsmath}
\usepackage{amssymb}
\usepackage{graphicx}
\usepackage{float}
\usepackage{epsfig}
\usepackage{color}
\usepackage{hhline}
\usepackage{pdfsync}
\newcommand{\ignore}[1]{}

\newcommand{\boxtheorem}{\hfill $\Box$}
\newcommand{\nit}[1]{{\it #1}}

\usepackage{times}

\usepackage{graphicx}

\usepackage{algpseudocode}
\usepackage{algorithm}
\usepackage{epstopdf}
\usepackage{enumerate}
\usepackage{times}

\newcounter{lemmaA-counter}
\newcounter{propositionA-counter}
{\vskip \abovedisplayskip \refstepcounter{lemmaA-counter}%
\noindent {\bf Lemma A.\arabic{lemmaA-counter}.}}%

{\vskip \abovedisplayskip \refstepcounter{propositionA-counter}%
\noindent {\bf Proposition A.\arabic{propositionA-counter}.}}%

\newcommand{\mc}[1]{\mathcal{ #1}}

\newcommand{\bcq}{BCQ}
\newcommand{\cq}{CQ}

\newcommand{\mf}[1]{\mathfrak{ #1}}

\newcommand{\fo}{FO}

\newcommand{\srep}{\nit{Rep}^{\sf S}(D,\Sigma)}

\title{\vspace*{-1cm} {\bf Measuring and Computing Database Inconsistency via Repairs}\vspace{-3mm}}
\author{{\bf Leopoldo Bertossi\vspace{-3mm}}\thanks{Member of the ``Millenium Institute for Foundational Research on Data" (IMFD, Chile). \ Email: bertossi@scs.carleton.ca. \ Research supported by NSERC Discovery Grant \#06148.}}\ignore{\thanks{ Carleton Univ., \
School of Computer Science, Canada. \
bertossi@scs.carleton.ca.}}

\institute{ Carleton University,  Ottawa, Canada \ \& \ $\mbox{Relational}^{\mbox{AI}}$\!, Inc.\vspace{-8mm}}

\begin{document}
\maketitle
\thispagestyle{empty}
\pagestyle{empty}

\ignore{
\vspace{-3mm}
\begin{figure}[ht]
\begin{center}
\vspace{-4mm}
 \includegraphics[width=9.5cm]{./fig/omdm} \vspace{-3mm}

 \caption{An \omd \ model with categorical relations, dimensional rules, and constraints}\label{fig:omdm} \vspace{-8mm}
\end{center}
\end{figure}
}
\begin{abstract}We propose a generic numerical measure of inconsistency of a database with respect to a set of integrity constraints. It is based on an abstract repair semantics. A particular inconsistency measure associated to cardinality-repairs is investigated; and we show that it can be computed via answer-set programs. \vspace{-6mm}
\end{abstract}



\vspace{3mm}Intuitively, a relational database may be more or less consistent than others databases for the same schema with the same integrity constraints (ICs). This comparison can be accomplished by assigning a {\em measure of inconsistency} to a database. The inconsistency degree of a database $D$ with respect to (wrt.) a set of ICs $\Sigma$ should depend on how complex it is to restore consistency; or more technically, on the class of {\em repairs} of $D$ wrt. $\Sigma$. For this we can apply concepts and results on database repairs  (cf. \cite{Bertossi2011} for a survey and references).
\ {\em Our stand on degrees of inconsistency  is that they depend on how consistency is restored, i.e. involving the admissible repair actions and how close we want stay to the instance at hand.} \ This short communication  shows preliminary research on possible ways to make these ideas concrete, by defining and analyzing  a measure of inconsistency of a relational database instance, and providing mechanisms for computing this measure  using {\em answer-set programming} (ASP) \cite{brewka}.

\vspace{-3mm}
\paragraph{\bf Database Repairs.} \ignore{\label{sec:reps}}  When a database instance $D$ does not satisfy its intended ICs, it is repaired, by deleting or inserting tuples from/into the database. An instance obtained in this way is a {\em repair} of $D$ if it satisfies the ICs and departs in a minimal way from $D$ \cite{Bertossi2011}.  In this work, just to fix ideas, we consider mostly ICs that can only be solved by tuple deletions, e.g. most prominently, {\em denial constraints} (DCs) and {\em functional dependencies} (FDs).   DCs are logical formulas of the form $\neg \exists \bar{x}(P_1(\bar{x}_1)\wedge \dots \wedge P_m(\bar{x}_m))$,
where  $\bar{x} = \bigcup \bar{x}_i$; and FDs are of the form  $\neg \exists \bar{x} (P(\bar{v},\bar{y}_1,z_1) \wedge P(\bar{v},\bar{y}_2,z_2) \wedge z_1 \neq z_2)$, with
$\bar{x} = \bar{y}_1 \cup \bar{y}_2 \cup \bar{v} \cup \{z_1, z_2\}$. \ We treat FDs as DCs. A database is {\em inconsistent} wrt. a set of ICs $\Sigma$ when $D$ does not satisfy $\Sigma$, denoted $D \not \models \Sigma$.

\vspace{-2mm}
\begin{example} \label{ex:rep} \  The DB $D = \{P(a), P(e), Q(a,b), R(a,c)\}$ is inconsistent with respect to the (set of) {\em denial constraints} (DCs) \ $\kappa_1\!: \ \neg \exists x \exists y (P(x) \wedge Q(x,y))$, and \
$\kappa_2\!: \ \neg \exists x \exists y (P(x) \wedge R(x,y))$. Here, \  $D \not \models \{\kappa_1, \kappa_2\}$.

\ignore{
\begin{multicols}{2}
\hspace*{-0.5cm}{\small
\begin{tabular}{c|c|}\hline
$P$&A\\ \hline
&a\\
&e\\ \hhline{~-}
\end{tabular}~~
\begin{tabular}{c|c|c|}\hline
$Q$&A&B\\ \hline
& a & b\\ \hhline{~--}
\end{tabular}~~
\begin{tabular}{c|c|c|}\hline
$R$&A&C\\ \hline
& a & c\\ \hhline{~--}
\end{tabular} }
{\begin{eqnarray*}
\psi_1\!: \ \neg \exists x \exists y (P(x) \wedge Q(x,y)),\\
\psi_2\!: \ \neg \exists x \exists y (P(x) \wedge R(x,y)).
\end{eqnarray*}}
\end{multicols}  }
A {\em subset-repair},  in short an {\em S-repair}, of $D$ wrt. the set of DCs is a $\subseteq$-maximal subset of $D$ that is consistent, i.e.  no proper superset is consistent. The following are
S-repairs: ${D_1 = \{P(e), Q(a,b), R(a,c)\}}$ and ${D_2 = \{P(e), P(a)\}}$. Under this repair semantics, both repairs are equally acceptable.
\ A {\em cardinality-repair},  in short a {\em C-repair}, is a maximum-cardinality S-repair.  $D_1$  is
the only C-repair. \boxtheorem
\end{example}

\vspace{-2mm}For an instance $D$ and a set $\Sigma$ of DCs, the sets of S-repairs and C-repairs are denoted with $\nit{Srep}(D,\Sigma)$ and $\nit{Crep}(D,\Sigma)$, resp. \ It holds: \ $\nit{Crep}(D,\Sigma) \subseteq \nit{Srep}(D,\Sigma)$.
More generally, for  a set $\Sigma$ of ICs, not necessarily DCs, they can be defined by (cf. \cite{Bertossi2011}): \
$\nit{Srep}(D,\Sigma) = \{D'~:~ D' \models \Sigma, \mbox{ and } D \bigtriangleup D' \mbox{ is minimal under set inclusion}\}$, and \
$\nit{Crep}(D,\Sigma) = \{D'~:~ D' \models \Sigma, \mbox{ and } D \bigtriangleup D' \mbox{ is minimal in cardinality}\}$.
\
Here, $D \bigtriangleup D'$ is the symmetric set difference $(D\smallsetminus D') \cup (D' \smallsetminus D)$.

\vspace{-3mm}
\paragraph{\bf Repair Semantics and Inconsistency Degrees.} \ In general terms, a {\em repair semantics} {\sf S} for  a schema $\mc{R}$ that includes a set $\Sigma$ of ICs assigns to each  instance $D$ for $\mc{R}$  (which may not satisfy $\Sigma$),  a class $\nit{Rep}^{\sf S}(D,\Sigma)$ of
{\sf S}{\em -repairs} of $D$ wrt. $\Sigma$, which are instances of $\mc{R}$ that satisfy $\Sigma$ and depart from $D$ according to some minimization criterion.
\ Several repair semantics have been considered in the literature, among them and beside those above, {\em prioritized repairs} \cite{stawo},  and {\em attribute-based repairs} that change attribute values by other data values, or by a null value, {\sf NULL}, as in SQL databases (cf. \cite{foiks18,Bertossi2011}).

According to our  take on how a database inconsistency degree depends on database repairs, we define the {\em inconsistency degree} of an instance $D$ wrt. a set of ICs $\Sigma$ in relation to a given repair semantics {\sf S}, as the  distance from $D$ to the class $\srep$: \vspace{-3mm}
\begin{equation}
\mbox{\nit{inc-deg}}^{\sf S}(D,\Sigma) := \nit{dist}(D,\srep). \label{eq:dist}
\end{equation}

\vspace{-1mm}This is an abstract measure that depends on {\sf S} and a chosen distance function \nit{dist}, from a world to a set of possible worlds. Under the assumption that any repair semantics should return $D$ when $D$ is consistent wrt. $\Sigma$ and
$\nit{dist}(D,\{D\}) = 0$, a consistent instance $D$ should have $0$ as inconsistency degree.\footnote{Abstract distances between two point-sets are investigated in \cite{eiterMannila}, with their computational properties. Our setting is a particular case.}

Notice that  the class $\srep$ might contain instances that are not sub-instances of $D$, for example, for different forms of {\em inclusion dependencies} (INDs) we may want to insert tuples;\footnote{For INDs repairs based only on tuple deletions can be considered \cite{chomicki}.} or  even under DCs, we may want to appeal to  attribute-based repairs. \ignore{For example, \cite{wijsen} investigates repairs of this kind; attribute values can be changed by other values in the data domain. In \cite{tkde,tplp} replacement on values by a null \`a la SQL (or at least that disallows joins and comparisons through it) are proposed and investigated, and similarly in \cite[sec. 7.4]{tocs}, to capture attribute-level causes.} {\em In the following we consider only repairs that are sub-instances of the given instance.} \ Still this leaves much room open for different kinds of repairs. For example, we may prefer to delete some tuples over others \cite{stawo}. Or, as in database causality \cite{suciu,tocs}, the database can be partitioned into {\em endogenous} and {\em exogenous} tuples, assuming we have more control on the former, or we trust more the latter; and we prefer {\em endogenous repairs} that delete preferably (only or preferably) endogenous tuples \cite{foiks18}.

\vspace{-3mm}
\paragraph{\bf An Inconsistency Measure.} \ Here we consider a concrete instantiation of (\ref{eq:dist}), and to fix ideas, only DCs. For them,  the repair semantics $\nit{Srep}(D,\Sigma)$ and $\nit{Crep}(D,\Sigma)$ are  particular cases of repair semantics
 {\sf S} where each $D' \in \srep$ is maximally contained in $D$. On this basis, we can define: \vspace{-3mm}
\begin{eqnarray}
\hspace*{-1mm}\mbox{\nit{inc-deg}}^{{\sf S},g_3\!}(D,\Sigma)  &:=&  \nit{dist}^{g_3\!}(D,\srep)  :=  \frac{|D| \! - \! \nit{max}\{ |D'| : D' \in \srep  \}}{|D|} \nonumber\\
&=& \frac{ \nit{min} \{|D \smallsetminus D'|~:~ D' \in \srep  \}}{|D|}, \hspace*{-1mm}\label{eq:distG3}
\end{eqnarray}

 \vspace{-3mm} \noindent inspired by distance $g_3$ in \cite{mannila} to measure the degree of violation of an FD by a database, whose satisfaction is restored through tuple deletions.\footnote{Other possible measures for single FDs and relationships between them can be found in \cite{mannila}.} This measure can be applied more generally as a ``quality measure", not only in relation to inconsistency, but also whenever  possibly several intended ``quality versions" of a dirty database exist, e.g. as determined by additional contextual information \cite{context}.

\vspace{-1mm}
\begin{example} (ex. \ref{ex:rep} cont.) \label{ex:rep2} \ignore{Consider again
$D = \{P(a), P(e), Q(a,b), R(a,c)\}$, which violates the set of DCs $\Sigma = \{\kappa_1, \kappa_2\}$.} Here, $\nit{Srep}(D,\Sigma) = \{D_1, D_2 \}$, and
$\nit{Crep}(D,\Sigma) = \{D_1 \}$\ignore{, with $D_1 = \{P(e),$ $ Q(a,b), R(a,c)\}$ and $D_2 = \{P(a), P(e)\}$}. They provide the inconsistency degrees:
\begin{eqnarray}
\mbox{\nit{inc-deg}}^{s,g_3\!}(D,\Sigma) &:=& \frac{4 - \nit{max}\{ |D'| ~:~D' \in \nit{Srep}(D,\Sigma)  \}}{4} =  \frac{4 -|D_1|}{4} = \frac{1}{4}, \label{eq:s}\\
\mbox{\nit{inc-deg}}^{c,g_3\!}(D,\Sigma) &:=& \frac{4 - \nit{max}\{ |D'| ~:~D' \in  \nit{Crep}(D,\Sigma) \}}{4} = \frac{4 - |D_1|}{4} = \frac{1}{4},  \label{eq:c}
\end{eqnarray}
 respectively. \boxtheorem
\end{example}

\vspace{-1mm}It holds  $\nit{Crep}(D,\Sigma) \subseteq \nit{Srep}(D,\Sigma)$, but $\nit{max}\{|D'|~:~D' \in \nit{Crep}(D,\Sigma)\}$ $ = \nit{max}\{|D'|~:~D' \in \nit{Srep}(D,\Sigma)\}$,  so it holds $\mbox{\nit{inc-deg}}^{s,g_3\!}(D,\Sigma) = \mbox{\nit{inc-deg}}^{c,g_3\!}(D,\Sigma)$. \ These measures always takes a value between $0$ and $1$. The former when $D$ is consistent (so it itself is its only repair).
 The measure takes the value $1$ only when $\srep = \emptyset$ \ (assuming that
 $\nit{max} \{|D'| ~:~ D' \in \emptyset\} = 0$), i.e. the database is {\em irreparable}, which is never the case for DCs and S-repairs: there is always an S-repair. However, it could be irreparable with different, but related repair semantics.
  For example, when we accept only endogenous repairs and none of them exists \cite{tocs}.

  \vspace{-2mm}
\begin{example} (ex. \ref{ex:rep2} cont.) Assume $D$ is partitioned into endogenous and exogenous tuples, say resp. \ $D = D^n \stackrel{.}{\cup} D^x$, with $D^n = \{ Q(a,b), R(a,c) \}$ and $D^x = \{ P(a), P(e)\}$. In this case, the endogenous-repair semantics that allows only a minimum number of deletions of endogenous tuples, defines the class of repairs: $\nit{Srep}^{c,n}(D,\Sigma) = \{D_2\}$, with $D_2$ as above. In this case,\footnote{For certain forms of {\em prioritized repairs}, such as endogenous repairs, the normalization coefficient $|D|$ might be unnecessarily large. In this particular case, it might be better to use $|D^n|$.} \ $\mbox{\nit{inc-deg}}^{c,n,g_3\!}(D,\Sigma) = \frac{4-2}{4} = \frac{1}{2}$.
\ Similarly, if now $D^n = \{ P(a), Q(a,b)  \}$ and $D^x = \{P(e), R(a,c)\}$, there are no endogenous repairs, and \ $\mbox{\nit{inc-deg}}^{c,n,g_3\!}(D,\Sigma) = 1$.
\boxtheorem
\end{example}

\vspace{-5mm}\paragraph{\bf ASP-Based Computation of the Inconsistency Measure.} \ We concentrate on measure $\mbox{\nit{inc-deg}}^{c,g_3\!}(D,\Sigma)$ \ (cf. (\ref{eq:c})).  More generally, we can start from  $\mbox{\nit{inc-deg}}^{s,g_3\!}(D,\Sigma)$, which can be computed through the maximum cardinality of an S-repair for $D$ wrt. $\Sigma$, or, equivalently, using the cardinality of a (actually, every) repair in $\nit{Crep}(D,\Sigma)$. In its turn, this can be done\footnote{This approach was followed in \cite{foiks18} to compute maximum {\em responsibility degrees} of database tuples as causes for violations of DCs, appealing to a causality-repair connection \cite{tocs}.} \ through compact specifications of repairs by means of ASPs. We just show an example.

\vspace{-2mm}\begin{example} (ex. \ref{ex:rep} cont.) For technical convenience, we insert global tuple-ids in $D$, i.e. $D=\{P(1,e), Q(2,a,b), R(3,a,c), P(4,a)\}$. \
It is possible to write an answer-set program, a {\em repair program}, $\Pi$ whose stable models $\mc{M}_1, \mc{M}_2$ are  correspondence with the repairs $D_1, D_2$, resp., namely
$\mc{M}_1 = \{P'(1,e,{\sf s}), Q'(2,a,b,{\sf s}), R'(3,a,c,{\sf s}),$ $ P'(4,a,{\sf d}) \}$ $ \cup \ D$ and $\mc{M}_2 = \{P'(1,e,{\sf s}), P'(4,a,{\sf s}), Q'(2,a,b,{\sf d}), R'(3,a,c,{\sf d}) \} \cup \ D$, where the primed predicates are nicknames for the original ones, and the annotations constants {\sf s}, {\sf d} indicate that the tuple stays or is deleted in/from the database, resp. \cite{foiks18,monica}

Now, to compute $\mbox{\nit{inc-deg}}^{c,g_3\!}(D,\Sigma)$, for the C-repair semantics, we can add rules to $\Pi$ to collect the {\em tids} of tuples deleted from the database: \ $\nit{Del}(t) \leftarrow R'(t,x,y,{\sf d})$, similarly for $Q'$ and $P'$. \ And next, a rule to count the deleted tuples, say: \
$\nit{NumDel}(n) \leftarrow \# \nit{count}\{t : \nit{Del}(t)\} = n$. \ For example, program $\Pi$ with the new rules added will see the original stable model $\mc{M}_1$ extended with the atoms $\nit{Del}(4), \nit{NumDel}(1)$. Similarly for $\mc{M}_2$.

Since the stable models of the program capture the S-repairs, i.e. $\subseteq$-maximal and consistent sub-instances of $D$, we can add to $\Pi$ {\em weak program constraints} \cite{dlv}, such as \ ``$:\sim  P(t,x), P'(t,x,{\sf d})$" (similarly for $R$ and $Q$). They have the effect of eliminating the models of the original program that do not violate them in a minimum way. More precisely, they make us keep  only the stable models of the original program that minimize the number of satisfactions of the constraint bodies. In our case, only the models (repairs) that minimize the number of tuple deletions are kept, i.e. models that correspond to C-repairs of $D$. In this example, only (the extended) $\mc{M}_1$ remains. The value for $\nit{NumDel}$ in any of them can be used to compute  $\mbox{\nit{inc-deg}}^{c,g_3\!}(D,\Sigma)$. There is no need to explicitly compute all stable models, their sizes, and compare them. This value can be obtained by means of the query, ``$\nit{NumDel}(x)?$", answered by the program  under the {\em brave semantics} (returning an answer from {\em some} stable model).
\boxtheorem \end{example}

\vspace{-5mm}
\paragraph{\bf Discussion.} \ There are many open issues, among them exploring other inconsistency measures, e.g. based on the {\em Jaccard distance} \cite{ullman}. Several measures have been considered in knowledge representation \cite{hunter,thimm,vanina}, mostly for the propositional case. \ It would be interesting to analyze the general properties of those measures that are closer to database applications, along the lines of
\cite{eiterMannila}; and their relationships. For each measure it becomes relevant to investigate the complexity of its computation, in particular, in data complexity (databases may have exponentially many repairs, in data \cite{Bertossi2011}).\footnote{Certain (or skeptical) reasoning with repair programs for DCs with weak constraints is $\Delta^P_2(\nit{log}(n))$-complete in data complexity, i.e. in the size of the database \cite{monica,dlv}.} Actually, it is possible to prove that computing $\mbox{\nit{inc-deg}}^{c,g_3\!}(D,\Sigma)$  is complete for the functional class $\nit{FP}^{\nit{NP(log(n))}}$ in data, and this both for sets $\Sigma$ of DCs and of FDs.

\vspace{-4mm}

\bibliographystyle{plain}

\end{document}